\documentclass[preprintnumbers,amsmath,amssymb,twocolumn,pra,showpacs,longbibliography]{revtex4-1}

\usepackage{mathtools}
\usepackage{algorithm}
\usepackage{setspace}
\usepackage{graphicx}
\usepackage{dcolumn}
\usepackage{amssymb}
\usepackage{color}
\usepackage{comment}
\usepackage{epstopdf}
\usepackage{bm}
\usepackage{enumerate}
\usepackage{fancyhdr,color}
\usepackage{verbatim}
\usepackage{amsmath,amsthm}
\usepackage{amsfonts}
\usepackage{hyperref}
\usepackage{xparse}
\usepackage{physics}
\usepackage{tikz}
\usepackage{soul}

\def\cE{\mathcal{E}}

\def\cI{\mathcal{I}}

\def\cO{\mathcal{O}}

\def\cU{\mathcal{U}}

\def\tr{\mathrm{tr}}
\def\Pr{\mathrm{Pr}}

\newtheorem{theorem}{Theorem}

\newtheorem{corollary}{Corollary}

\def\one{{\mathchoice {\rm 1\mskip-4mu l} {\rm 1\mskip-4mu l} {\rm
1\mskip-4.5mu l} {\rm 1\mskip-5mu l}}}
\def\dqc1{\textsc{DQC1}}

\newcommand{\sket}[1]{\,| #1 \rangle \,}  
\newcommand{\sbra}[1]{\langle #1|}

\begin{document}
    \title{Complexity of quantum state verification in the quantum linear systems problem}

\author{
Rolando D. Somma}
\affiliation{Theoretical Division, Los Alamos National Laboratory, Los Alamos, NM 87545, USA.}

\author{Yi\u{g}it Suba\c{s}\i}
\affiliation{Computer, Computational, and Statistical Sciences Division, Los Alamos National Laboratory, Los Alamos, NM 87545, USA.}

\date{\today}

\begin{abstract}
We analyze the complexity of quantum state verification
in the context of solving systems of linear equations of the form $A \vec x = \vec b$. 
We show that any quantum operation that verifies whether 
a given quantum state is within a constant distance from the
solution of the quantum linear systems problem
requires $q=\Omega(\kappa)$ uses of a unitary that prepares
a quantum state $\ket b$, proportional to $\vec b$, and its inverse in the worst case. Here, $\kappa$
is the condition number of the matrix $A$.
For typical instances, we show that $q=\Omega(\sqrt \kappa)$ with high probability. These lower bounds are almost
achieved if quantum state verification is performed using
known quantum algorithms for the quantum linear systems
problem. We also analyze the number of copies of $\ket b$ required by verification procedures of the prepare and measure type. In this case, the lower bounds are quadratically worse, being $\Omega(\kappa^2)$ in the worst case and $\Omega(\kappa)$ in typical instances with high probability. We discuss the implications of our results to known variational and related approaches to this problem, where state preparation, gate, and measurement errors will need to decrease rapidly with $\kappa$ for worst-case and typical instances if error correction is not used, and present some open problems.
\end{abstract}

\maketitle    
    
\section{Introduction}
\label{sec:intro}

Quantum computers may solve some 
problems that appear to be beyond reach of classical computers. 
Many examples of quantum speedups now exist, 
from the former result of P. Shor on the factoring problem~\cite{Sho97}, to optimization~\cite{SBB08},
the simulation of quantum systems~\cite{GAN14}, and beyond. As quantum technologies advance~\cite{LFL10},
so is the field of theoretical quantum computing, fueling the quest for new and fast quantum algorithms.

Along this quest, there has been interest in quantum algorithms for linear algebra, 
in particular for a problem related to solving systems of linear equations
of the form $A \vec x = \vec b$. This problem, which we refer to as the quantum linear systems problem (QLSP), was  introduced
in Ref.~\cite{HHL09}. There, a quantum algorithm was given -- the HHL algorithm --  and its complexity was shown to be polylogarithmic in $N$, the dimension of the matrix $A$, under some assumptions. Due to the potential for an exponential quantum speedup and the relevance of systems of linear equations in science, the results of Ref.~\cite{HHL09} sparked 
further interest for improved versions of the HHL algorithm. For example, Refs.~\cite{Amb12,CKS17,SSO19,AL19} provide quantum algorithms for the QLSP with provable runtimes that are almost linear in $\kappa$, the condition number of $A$.
These algorithms run faster than HHL, whose complexity is quadratic in $\kappa$, and are almost optimal.

More recently, quantum algorithms
for the QLSP inspired by variational and related approximation approaches
were given in Refs.~\cite{AL19,HBR19,BLC19,XSE19}. In a variational approach, the algorithm is designed via an optimization loop
that requires preparing multiple copies of a parametrized quantum state, measuring a cost function, and using the measurement information to update the parameters for the next round of state preparations. This process is repeated until the cost function is minimized. Variational approaches open the possibility of preparing quantum states and solving certain problems with less complexity than the best-known methods, e.g., shorter circuit depths, or less number of qubits~(cf.~\cite{PMS14,FGG14,khatri2019quantum}),
making them attractive to near-term applications. Similar arguments may also hold for other quantum algorithms, such as those formulated in the quantum adiabatic model~\cite{FGGS00}. In this case, one may attempt
to execute the evolution in less time than known upper bounds, with the potential of solving a problem with improved complexity (cf.~\cite{SNK12,CFC14}).

Like all heuristics, the actual runtime of these approaches
may be unknown a priori and the algorithms stop when a particular
criterion is satisfied. 
For the QLSP, this requires verifying that the prepared quantum state is indeed sufficiently close to the desired one. This {\em quantum state verification} (QSV) step requires additional resources that need to be accounted for
when determining the overall complexity of such approaches. A question then arises: Can QSV be performed with {\em low} complexity? 

In this paper, we answer this question in the negative. This is striking because 
the solution to many computational problems 
can be verified in significantly less time than producing the solution itself, such as for NP-complete problems~\cite{VL90}, but  this is not the case for the QLSP.
In particular, we show that the complexity
for QSV in the QLSP is $\Omega(\kappa)$ in the worst case. More precisely,
if a quantum state $\sket b$ that encodes the vector $\vec b$ can only be accessed via its preparation unitary $U_b^{\;}$, then the number of uses of $U_b^{\;}$, $U_b^{-1}$, and their controlled versions $cU_b^{\pm 1}$ needed for QSV is $\Omega(\kappa)$ in the worst case. For typical instances of the QLSP, these unitaries must be implemented $\Omega(\sqrt \kappa)$ times with high probability. As $\kappa$ can grow rapidly with the problem size, perhaps scaling with the dimension of $A$, which is the case
for many applications~\cite{Ede88},
QSV in the QLSP can be expensive.

Our main result is a generic lower bound for the complexity of QSV that applies to any instance of the QLSP.
We also prove that optimal QSV, in terms of uses of $cU_b^{\pm 1}$ (or $U_b^{\pm1}$), can be achieved using a known
quantum algorithm for the QLSP, such as the
HHL algorithm~\cite{HHL09}.
One can run this algorithm to solve the QLSP and prepare
a quantum state $\ket x$ proportional to $\vec x$,
and then use the well-known swap test 
to verify if a given quantum state is close to $\ket x$~\cite{BCW01}. Although the HHL algorithm is not optimal~\cite{Amb12,CKS17,SSO19,AL19}, it turns out that is almost optimal in terms of uses of $cU_b^{\pm 1}$.

We also analyze a restricted class of QSV procedures
of the prepare and measure type. In this case,
we are given $q \ge 1$ copies of the quantum
state $\ket b$ and arbitrarily many copies of the state to be verified, and the QSV
procedure only involves a joint measurement of all quantum systems. We prove that $q=\Omega(\kappa^2)$ in the worst case and $q=\Omega(\kappa)$ for typical instances  
of the QLSP with high probability. These lower bounds
are quadratically worse than those for general QSV procedures.

Our results place limitations
for approaches to the QLSP that require a QSV step.
If QSV is performed via
the computation of a simple cost function that does not exploit the structure of $U_b$, such as in known variational approaches,
then the number of state preparations and projective measurements must increase rapidly (i.e., polynomially) with $\kappa$ for worst-case and typical instances. Thus, to avoid error correction,
state-preparation, gate, and measurement errors
 need to decrease rapidly with $\kappa$, which is unrealistic. Nevertheless, our lower bounds on the complexity of QSV, as well as those for solving the QLSP~\cite{HHL09}, may be bypassed if the structure of $U_b^{\;}$ can be exploited, opening the possibility
for faster approaches.

The rest of the paper is organized as follows.
In Sec.~\ref{sec:QLSP} we describe the QLSP in detail.
In Sec.~\ref{sec:QSV} we introduce the QSV problem
for the QLSP and present our main results, focusing on worst-case and typical instances. In Sec.~\ref{sec:optimalQSV} we describe an almost optimal
QSV procedure based on the HHL algorithm. In Sec.~\ref{sec:PMQSV} we analyze the complexity of QSV procedures of the prepare and measure type.
In Sec.~\ref{sec:implications} we give more details on the implications of our results, the limitations of variational approaches to the QLSP, and present some open problems. We provide further conclusions in Sec.~\ref{sec:conc}. Detailed proofs
of our main results are in the Appendices.

\section{The QLSP}    
\label{sec:QLSP}

We introduce the QLSP following Refs.~\cite{HHL09,Amb12,CKS17,SSO19}.
We are given an $N \times N$ Hermitian and non-singular matrix $A$, $N \ge 2$, a vector $\vec b = (b_0,b_1,\ldots, b_{N-1})^T$, and a precision parameter $\epsilon>0$. 
The matrix has spectral norm $\|A \| = 1$ and its condition number, which is the ratio between the absolute largest and smallest eigenvalues
of $A$, is $\kappa < \infty$.
We define
\begin{align}
    \ket x : = \frac {\sum_{j=0}^{N-1} x_j \ket j}
    {\| \sum_{j=0}^{N-1} x_j \ket j\|} \;,
\end{align}
which is a unit (pure) quantum state proportional to the solution of the system $A \vec x = \vec b$,
where $\vec x = (x_0,x_1,\ldots,x_{N-1})^T$ is the solution. In general, we write $\| \ket a \|$ for the Euclidean norm of a quantum state $\ket a$. If $\ket b$ is a quantum state proportional to $\vec b$, then $\ket x = A^{-1}\ket b/ \| A^{-1}\ket b \|$. 

The goal in the QLSP is to prepare a (possibly mixed) quantum state $\rho$ 
that satisfies
\begin{align}
\label{eq:solution}
   D_{\rho,x}:= \frac 1 2 \| \rho - \ketbra x \|_{\tr} \le \epsilon \; ,
\end{align}
where $\| X \|_{\tr}= \tr(\sqrt {XX^\dagger})$ is the trace norm
of  $X$ and $D_{\rho,x}$ is the trace distance.

Equation~\eqref{eq:solution} implies that no experiment can distinguish $\rho$ from $\ket x$ with probability greater than $\epsilon$ in a single shot~\cite{Ba09}. Additionally, the expectation of an operator in $\rho$ differs from that in $\ket x$ by an amount that is, at most, proportional to $\epsilon$.
We assume $N=2^n$ without loss of generality, so that $\rho$, $\ket x$, and $\ket b$ are $n$-qubit states.

For the QLSP, we need to specify $\vec b$ and $A$ in some way. Like
known quantum algorithms, here we assume access to a unitary $U_b^{\;}$ and its inverse $U_b^{-1}$, such that $\ket b = U_b^{\;}\ket 0$.
The state $\ket 0$ is some simple state of $n$ qubits, such as the all-zero state. We further assume access to the controlled versions of these unitaries, $cU_b^{\pm 1}$, which are more powerful and implement $U_b^{\pm 1}$ only when the state of a control qubit is $\ket 1$ and do nothing otherwise.
For the matrix $A$, one may assume access to a procedure $U_A$
that computes the positions and values of the nonzero entries of $A$, but other access models
could also be of interest.
The operations $cU_b^{\pm 1}$ are treated as ``black boxes'', and no assumptions
are made on the inner workings of such unitaries.

Our results regarding the complexity of QSV are  lower bounds on the uses of $U_b^{\pm 1}$ only or, more generally, $cU_b^{\pm 1}$. These bounds often have implications on the gate requirements for certain QSV approaches as well.
Analyzing other complexities, such as the number of uses of $U_A$ required for QSV, may also be of interest. However, it is reasonable to expect that early applications of the QLSP will not use $U_A$ but rather some other way of implementing $A^{-1}$, where such results would not directly apply.

\section{QSV and main results}
\label{sec:QSV}

We seek to certify whether $D_{\rho,x} \le 1/8$ or $D_{\rho,x} > 1/2$ for a given quantum state $\rho$. We choose these limits for simplicity, as these suffice
for our purposes, but generalization
to arbitrary gap between the limits is simple. 
In the QSV problem the goal is 
to construct a protocol that, on inputs $A$ and $\vec b$, returns a quantum operation for QSV.
We write $\cE$
for the quantum operation, which is a completely-positive and trace preserving (CPTP) map,  and note that $\cE$
can access $\vec b$  via the action of $cU_b^{\pm 1}$ only. The quantum operation $\cE$ takes arbitrary many copies of $\rho$ as input and outputs a random bit $r$ as follows:
\begin{align}
\label{eq:QSVcond}
    \Pr (r=1)\ \left \{ \begin{matrix} & \ge 2/3 \ \ \text{if} \ \ D_{\rho,x} \le 1/8 \; , \cr
   & \le 1/3 \ \ \text{if} \ \ D_{\rho,x} > 1/2 \;. \end{matrix} \right.
\end{align}
When $r=1$, we claim that $\rho$ ``passed the test'' or that $\cE$ ``accepted'' $\rho$, implying that $\rho$
is likely to be {\em close} to $\ket x$.
When $r=0$, we claim that $\rho$  ``failed the test'' or that $\cE$ ``rejected'' $\rho$, implying that $\rho$ is likely to be {\em far} from $\ket x$.
One can amplify the probabilities
of passing or failing the test from $2/3$ to near 1 in either case by repetition and taking the median of the outcomes.

For a different instance specified by the same matrix $A$ but different vector $\vec b'$, the QSV protocol returns a quantum operation $\cE'$ that is different from $\cE$. The two operations differ only in the state-preparation unitaries, and $\cE'$ can be obtained from $\cE$ by replacing $cU_b^{\pm 1} \rightarrow cU_{b'}^{\pm 1}$.

\begin{figure}[htb]
    \includegraphics[width=8.65cm]{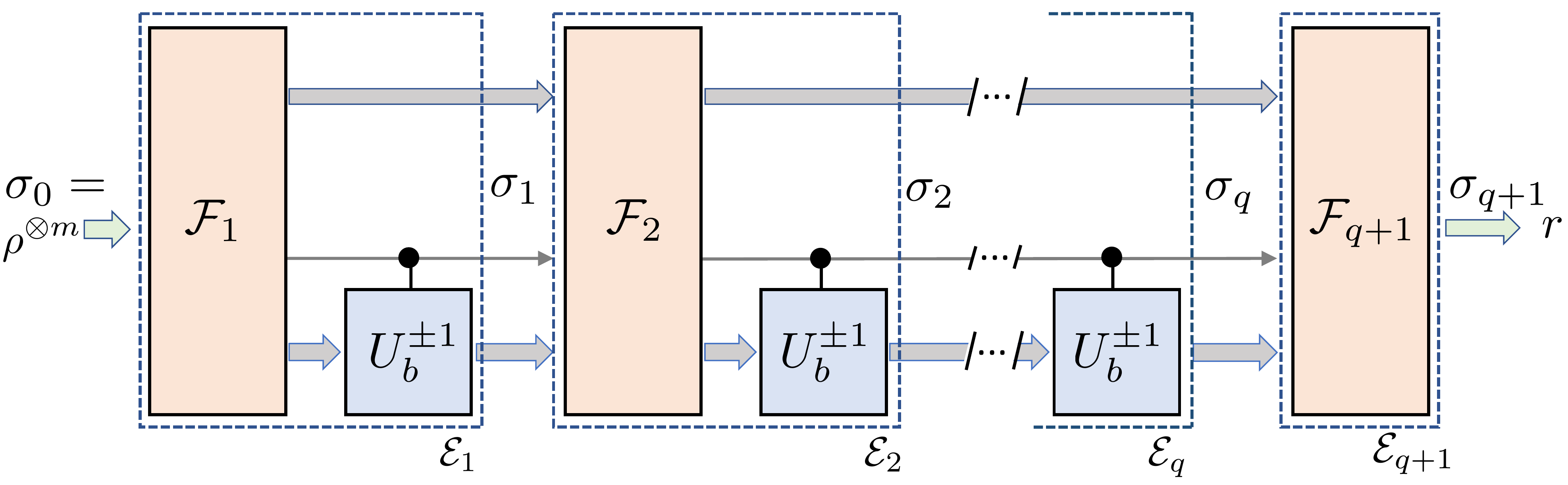}
     \caption{The quantum operation $\cE$.
     Arrows denote the states $\sigma_j$ output by the quantum operations $\cE_j$ and used as the input to the following $\cE_{j+1}$.
     The input state $\sigma_0$ contains $m$ copies of a state $\rho$. The output state $\sigma_{q+1}$ contains the bit $r$.}
    \label{fig:quantumprocess}
\end{figure}

In general, $\cE$ will contain measurements
and unitaries, including
$U_b^{\pm 1}$ and $cU_b^{\pm 1}$, and can be described as in Fig.~\ref{fig:quantumprocess} without loss of generality.
In this case, $\cE=\cE_{q+1} \circ \dots \circ \cE_1$ is a composition of $q+1 \ge 1$ quantum operations $\cE_j$.
For $j \le q$, these are
\begin{align}
    \cE_j :=  {\cal U}_b^{s_j} \circ {\cal F}_j  \; ,
\end{align}
where the ${\cal F}_j$'s are quantum operations that do not use  $cU_b^{ \pm 1}$ (or $U_b^{\pm1}$), ${\cal U}_b^{s_j}$ is the quantum operation that implements the unitary $cU_b^{s_j}$ on part of the register output by ${\cal F}_{j-1}$,
and $s_j = \pm 1$.  
The input to ${\cal F}_1$ (and $\cE$)
is a state $\sigma_0$ composed of $m \ge 1$ copies of a quantum state $\rho$. The output
of ${\cal F}_{q+1}$ (and $\cE$) contains the bit $r$.
Note that, if $\cE$ initially used unitaries $U_b^{\pm 1}$ that were not controlled, or if these unitaries were controlled on the state of a classical bit, these can still be thought as $cU_b^{\pm 1}$'s with a proper state for the control qubit (e.g., $\ket 1$). We then measure the complexity of a generic QSV procedure by the number of $cU_b^{\pm 1}$ required for its implementation.

As defined, $q$ is the maximum number of $cU_b^{\pm 1}$
needed to implement $\cE$. Nevertheless, the actual
number of such unitaries implemented on any one execution
of $\cE$, $q_{A,b}$, may be random and less than $q$; only $q$ such unitaries are needed in the worst case.
For example, the operation could stop once certain criterion
is met, such as a (random) measurement outcome. 
Our main result places a lower bound on $q_{A,b}$
that must be satisfied with constant probability by
any quantum operation for QSV, for any $m \ge 1$, and for any instance of the QLSP:
\begin{theorem}
\label{thm:main}
Consider any instance of the QLSP, specified by $A$ and $\vec b$,
and any protocol for QSV
 as above.
Then, for all quantum states $\rho$ that satisfy $D_{\rho,x}\le 1/8$, the number 
of $cU_b^{\pm 1}$'s required to implement $\cE$ on input $\sigma_0=\rho^{\otimes m}$ satisfies
\begin{align}
\label{eq:mainresult}
{\rm Pr}\left(q_{A,b} >  \frac 1 {13} \; \frac \kappa { \|A^{-1} \ket b\| }\right) \ge \frac{1}{6} \; .
\end{align}
\end{theorem}

The proof of Thm.~\ref{thm:main} is in Appendix~\ref{app:mainproof}. The basic 
idea is related to that of Ref.~\cite{BBBV97} for proving the lower bound on quantum search
 but is different in that the result is probabilistic and
the lower bound depends on the problem instance.
It works as follows: For any $\vec b$ (and fixed $A$), it is possible
to construct another instance specified by $\vec b'$,
where the solutions to the corresponding QLSPs satisfy
$D_{x,x'}:=\frac 1 2 \|\ketbra x - \ketbra{x'}\|_{\tr}>5/8$. Thus, $D_{\rho,x'}>1/2$ and $\cE$ must accept $\rho$ with large probability ($\ge 2/3$) while $\cE'$, which is the QSV operation that uses $cU_{b'}^{\pm 1}$, must reject it with large probability ($\ge 2/3$). Otherwise Eq.~\eqref{eq:QSVcond} is not satisfied. Simultaneously, the controlled state-preparation unitaries for these instances are shown to satisfy $\|cU_b^{\pm 1} - cU_{b'}^{\pm 1}\|=\cO(\|A^{-1}\ket b\|/\kappa)$. As $\cE$ and $\cE'$
differ only in these unitaries (i.e., the operations ${\cal F}_j$ are the same), the only way to distinguish among these two operations, or produce
a constant change in $\Pr(r)$ on input $\sigma_0$, is if the unitaries are
used $\Omega(\kappa/\|A^{-1}\ket b\|)$ times, with constant probability.

The argument behind Thm.~\ref{thm:main} thus provides a relation
between the complexity of QSV and the changes in the solution of the QLSP under perturbations to the initial state $\ket b$. 
This {\em susceptibility} is indeed quantified by $\kappa/\|A^{-1}\ket b\|$ as seen from the following examples.

\subsection{Worst-case instances}
\label{sec:worstcase}

\begin{corollary}
\label{cor:worstcase}
There exist instances of the QLSP such that for all quantum states $\rho$ that satisfy $D_{\rho,x}\le 1/8$, the number of $cU_b^{\pm 1}$'s required to implement $\cE$
on input $\sigma_0=\rho^{\otimes m}$ satisfies
\begin{align}
    {\rm Pr} \left( q_{A,b} > \frac 1 {13} \; \kappa\right) \ge \frac{1}{6} \; .
\end{align}
\end{corollary}

This result is a direct consequence of Thm.~\ref{thm:main},
obtained by replacing $\|A^{-1} \ket b\| \rightarrow 1$, 
which occurs when $\ket b$ is supported on eigenstates of $A$
of eigenvalue $\pm 1$ only. (Note that, in general, $\|A^{-1}\ket b\| \ge 1$.) For these instances, the susceptibility is large: a {\em small} change in $\ket b$ can result in a {\em big} change in $\ket x$. For example, if $\ket b=\ket x = \ket 1$ and we replace $\ket b \rightarrow \ket{b'} \propto \ket b +  (1/\kappa) \sket{(1/\kappa)}$, where $\ket 1$ and $\sket{(1/\kappa)}$
are eigenstates of $A$ of eigenvalue $1$ and $1/\kappa$, respectively, the solution to the new QLSP is $\sket{x'}=(\ket 1 + \sket{(1/\kappa)})/\sqrt 2$.
These states satisfy
$D_{x,x'}=\sqrt{1/2}>1/2$. At the same time, the states $\ket b$ and $\sket{b'}$ can be prepared with two unitaries $U_b^{\;}$ and $U_{b'}^{\;}$ that satisfy $\|U_b^{\pm 1} - U_{b'}^{\pm 1}\|=\|cU_b^{\pm 1} - cU_{b'}^{\pm 1}\|=\cO(1/\kappa)$.
Following Thm.~\ref{thm:main}, 
the number of $cU_b^{\pm 1}$'s needed to implement the QSV operation $\cE$, on input $\sigma_0$, is $\Omega(\kappa)$ with constant probability ($\ge 1/6$).

\subsection{Typical instances}
\label{sec:typical}

The quantity
$\|A^{-1} \ket b\|$ can take any value in $[1,\kappa]$ providing a wide range of lower bounds when $\kappa \gg 1$.
It is important to determine $\|A^{-1} \ket b\|$ in typical instances of the QLSP since
a lower bound on the complexity of QSV in such instances may differ from those in the worst or best cases.
To this end, we consider instances where the eigenvalues of $A$ are sampled from the uniform distribution $\rm{unif}\{[-1,-1/\kappa] \cup [1/\kappa,1]\}$ and the amplitudes in the spectral decomposition of $\ket b$ are sampled from the so-called Porter-Thomas distribution (and renormalized)~\cite{PT56}. 
This scenario resembles the one where the initial state $\ket b$
is prepared by a random quantum circuit~\cite{BIS18,Aru19}.
We obtain:
\begin{theorem}
\label{thm:typical}
Consider a random instance of the QLSP as described above.
Then, there exists a constant $c>0$ such that
\begin{align}
    \Pr ( \| A^{-1} \ket b \| \notin [\sqrt{ \kappa/2},\sqrt{3 \kappa /2}] ) \le 4 e^{-c N/\kappa} \;.
\end{align}
\end{theorem}

The proof of Thm.~\ref{thm:typical} is in Appendix~\ref{app:typical}. In the asymptotic limit where $N \gg \kappa$, we obtain that $\| A^{-1} \ket b \|=\Theta(\sqrt \kappa)$ with overwhelming probability.
This implies:
\begin{corollary}
\label{cor:typical}
Consider a random instance of the QLSP as described above.
Then, there exists a constant $c>0$ such that, for all quantum states $\rho$ that satisfy $D_{\rho,x}\le 1/8$,
the number of $cU_b^{\pm 1}$ required to implement $\cE$
on input $\sigma_0=\rho^{\otimes m}$ satisfies
\begin{align}
    {\rm Pr}\left( q_{A,b} > \frac 1 {16} \sqrt \kappa \right) \ge \frac{1-4e^{-cN/\kappa}}{6} \; .
\end{align}
\end{corollary}

Corollary~\ref{cor:typical} is a direct consequence of Thms.~\ref{thm:main} and~\ref{thm:typical}, where we replaced $\|A^{-1}\ket b\| \rightarrow \sqrt{3 \kappa /2}$ and bounded the joint probability by the product of $(1-4 e^{-cN/\kappa})$, which is a lower bound on the probability that $\|A^{-1}\ket b\| \le \sqrt{3 \kappa/2}$, and $1/6$, which is the lower bound on the probability in Thm.~\ref{thm:main} that applies to any instance.
Thus, for typical instances of the QLSP and $N/\kappa=\Omega(1)$, the complexity of QSV is
$\Omega(\sqrt \kappa)$ with constant probability.

\section{Optimal QSV procedure}
\label{sec:optimalQSV}

According to Thm.~\ref{thm:main}, any quantum operation
for QSV in the QLSP requires $\Omega(\kappa/\|A^{-1}\ket b\|)$
uses of $cU_b^{\pm 1}$ in expectation. An optimal QSV procedure
is one that achieves this bound. In this section, we show that the former HHL algorithm can provide an almost optimal procedure for QSV, despite not being an optimal algorithm for solving the QLSP: the number of calls to the procedure $U_A$ is quadratic, rather than linear, in $\kappa$ and polynomial in the inverse of a precision parameter.  Other known quantum algorithms for the QLSP could also be used for optimal QSV and require less $U_A$'s~\cite{Amb12,CKS17}. 

We use the HHL algorithm to prepare a state $\rho_x$ that is close to $\ket x$. Then, we implement the swap test~\cite{BCW01} to gain information about the distance between $\rho_x$ and $\rho$, and thus between $\ket x$ and $\rho$.
In Appendix ~\ref{app:HHLQSV} we show that in order to satisfy Eq.~\eqref{eq:QSVcond} it suffices to require $D_{\rho_x,x}=\Omega(1)$ and to implement the HHL algorithm and the swap test a constant number of times.

The HHL algorithm prepares $\rho_x$ using the unitaries $cU_b^{\pm 1}$ a number of times that is $\cO(\kappa/\|A^{-1}\ket b\|)$ in expectation. To achieve this, the HHL algorithm first applies an approximation of $A^{-1}/\kappa$ to $\ket b$ using quantum phase estimation and then uses amplitude amplification to amplify 
the probability of observing $\ket x \propto (A^{-1}/\kappa) \ket b$. The expected number of amplitude amplification rounds is $\cO(\kappa/\|A^{-1}\ket b\|)$, the inverse of the norm of $(A^{-1}/\kappa)\ket b$, if we follow  Ref.~\cite{BHMT02}. 


\section{Prepare and measure (PM) approaches}
\label{sec:PMQSV}

The results in Sec.~\ref{sec:QSV} consider  quantum operations for QSV that assume access to the unitaries $cU_b^{\pm 1}$. In contrast, prepare and measure (PM) approaches to QSV do not make this assumption. In a PM approach we are only allowed to prepare multiple copies of $\ket b$, multiple copies of $\rho$, and perform a joint measurement of all systems that produces the bit $r$ according to Eq.~\eqref{eq:QSVcond}. The joint measurement only involves operations that do not depend on $\vec b$, but they may depend on $A$.

\begin{figure}[htb]
    \includegraphics[width=8.5cm]{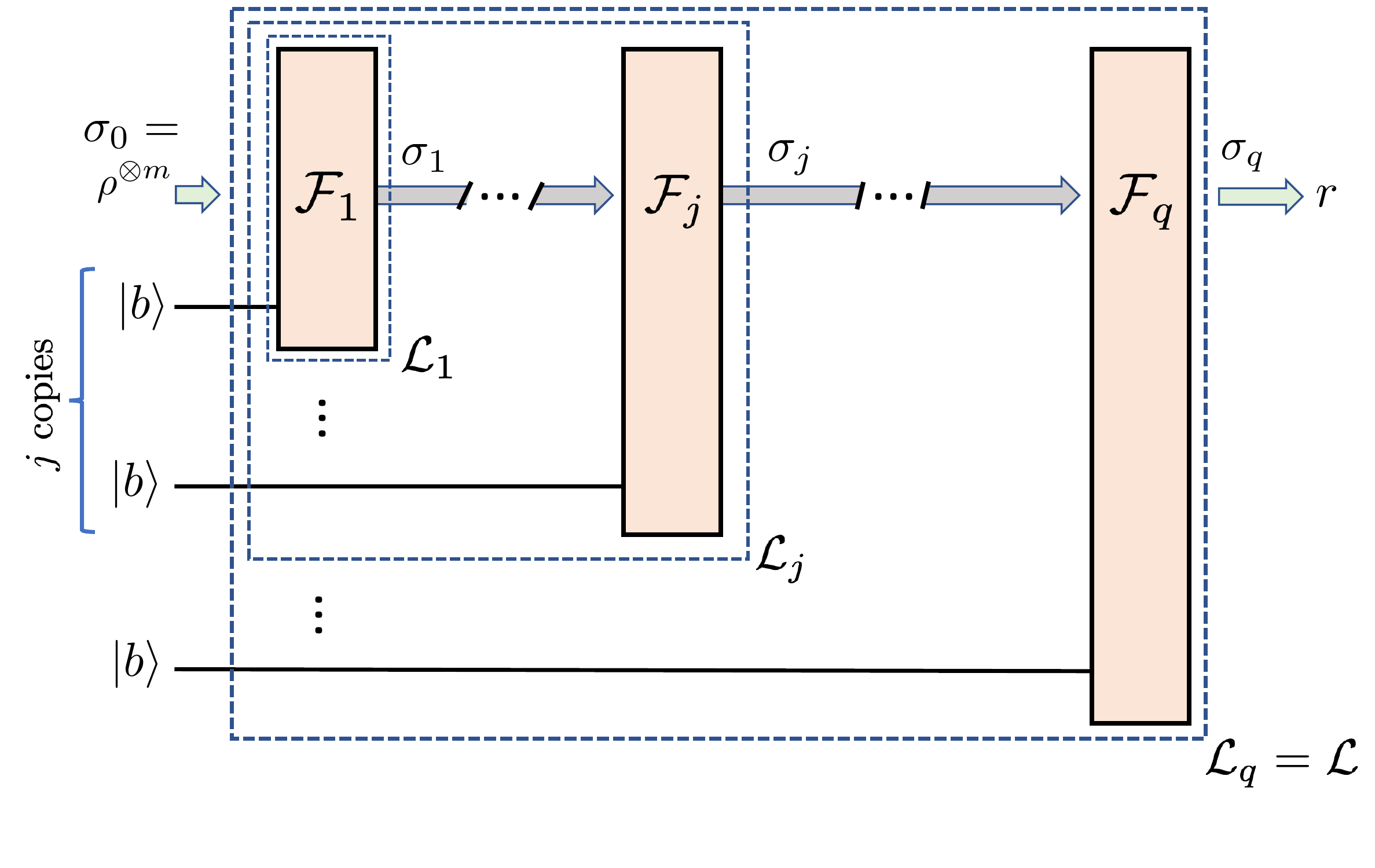}
     \caption{The quantum operation $\cal L$ of the PM type returned by the QSV protocol on input $A$. Arrows denote the states $\sigma_j$ output by the quantum operations ${\cal F}_j$ and used as the input to the following ${\cal F}_{j+1}$, together with a fresh copy of $\ket{b}$.
     The input state to $\cal L$ contains $m$ copies of $\rho$ and $q$ copies of $\ket{b}$. The output state contains the bit $r$.}
    \label{fig:quantumprocess2}
\end{figure}

Any PM protocol for QSV returns then a quantum operation ${\cal L}$ that depends on $A$.
This operation can be described
as in Fig.~\ref{fig:quantumprocess2} without loss of generality.
It
is a sequence of $q \ge 1$ operations ${\cal F}_j$, where each ${\cal F}_j$ is independent of $\vec b$ and takes as input the state output by ${\cal F}_{j-1}$,
together with a fresh copy of $\ket b$. The input to ${\cal F}_1$ is the state $\sigma_0=\rho^{\otimes m}$, $m \ge 1$, and a copy of $\ket{b}$. The output of ${\cal F}_{q}$ (and ${\cal L}$)
contains the bit $r$. We measure the complexity
of a PM approach to QSV by the number of copies of $\ket b$
required in the input  of ${\cal L}$.

As defined, $q$ is the maximum number of copies of $\ket b$
used by ${\cal L}$. Nevertheless, the actual number of such states used in any one execution of ${\cal L}$, $q_{A,b}$, may be random and less than $q$; only $q$ such states are required in the worst case.
The following result is the analogue of Thm.~\ref{thm:main} for a PM approach to QSV. It places a lower bound on $q_{A,b}$
that must be satisfied with constant probability by any quantum operation for QSV of the PM type, for any $m \ge 1$, and for any instance of the QLSP:
\begin{theorem}
\label{thm:PM}
Consider any instance of the QLSP specified by $A$ and $\vec b$, and any protocol for QSV of the PM type as above. Then, for all quantum states $\rho$ that satisfy $D_{\rho,x}\le 1/8$, the number 
of copies of $\ket{b}$ required by ${\cal L}$  for $\sigma_0=\rho^{\otimes m}$ satisfies
\begin{align}
\label{eq:PMresult}
{\rm Pr} \left(q_{A,b} > \frac 1 {{150}} \frac{\kappa^2}{ \|A^{-1} \ket b\|^2 } \right) \ge \frac{1}{6} \; .
\end{align}
\end{theorem}

The detailed proof is contained in Appendix~\ref{app:PMproof} and the basic 
idea is similar to that of Thm.~\ref{thm:main}, in that we consider two instances of the QLSP, where $A$ is fixed but $\vec b \ne \vec b'$. In this case, the quantum operation ${\cal L}$ is fixed but it is the input state
what changes when we replace $\ket b \rightarrow \sket{b'}$.
The number of copies of this state needs to be sufficiently large to produce a constant change in $\Pr(r)$, according to Eq.~\eqref{eq:QSVcond}, setting the lower bound in Thm.~\ref{thm:PM}.
The scaling in Eq.~\eqref{eq:PMresult} is quadratically worse than that obtained when one has direct access to both unitaries $cU_b^{\pm 1}$. This is because
the trace distance between $q$ copies of $\ket b$
and $q$ copies of $\sket{b'}$ scales as $\sqrt q$
rather than linear in $q$.

Following Secs.~\ref{sec:worstcase} and~\ref{sec:typical}, $\Omega(\kappa^2)$
copies of $\ket b$ will be needed for the PM approach in the worst case and $\Omega(\kappa)$ in the typical case, with constant probability.

\section{Implications and open problems}
\label{sec:implications}

We analyze some implications of Thms.~\ref{thm:main} and~\ref{thm:PM} in more detail and provide some open problems, which aim at bypassing our lower bounds. First, we note that the lower bounds are independent of $m$. Even if we had access to a full classical description of $\rho$, which could be obtained via quantum state tomography using $m \gg 1$ copies, the QSV procedures would still need $\Omega(\kappa/\|A^{-1} \ket b \|)$ uses of $cU_b^{\pm 1}$
or $\Omega((\kappa/\|A^{-1} \ket b \|)^2)$ copies of $\ket b$ with constant probability, respectively.  Our results also suggest that we must know (or compute) $\kappa$ beforehand, to be confident that a given QSV procedure works. For example, if $\kappa \gg 1$
but a given QSV procedure involves a few ($\ll \kappa$) unitaries $cU_b^{\pm 1}$ or a few copies of $\ket b$, then Thms.~\ref{thm:main} and~\ref{thm:PM} imply that such a procedure cannot produce the bit $r$ that satisfies Eq.~\eqref{eq:QSVcond}.

Additionally, known variational and related approximation algorithms for the QLSP also require a  (weaker) form of
QSV~\cite{AL19,HBR19,BLC19,XSE19}. To this end, these algorithms evaluate
a cost function such as $C(\rho)=\Tr(\rho H)$, which is the expectation of an observable $H \ge 0$ in $\rho$. In general, $C \ge 0$, and $C=0$ only when the state is the solution to the QLSP; that is, $C(\ketbra x)=0$. The cost function can be used
to detect some states that are close to $\ket x$. For example, one can set a threshold $C_{\min}>0$ such that, if $C(\rho) \le C_{\min}$ then $D_{\rho,x} \le 1/8$. As $C$ is estimated within given confidence, we can set this to be, at least, $2/3$. 

This weaker form of QSV requires a protocol that,
on input $A$ and $\vec b$, provides a quantum operation $\cE$ with the following properties for the bit $r$:
\begin{align}
\label{eq:weakQSVcond}
    \Pr (r=1)\ \left \{ \begin{matrix} & \ge 2/3 & \text{if} \ \ C(\rho) \le C_{\min} \; , \cr
   & \le 1/3 & \text{if} \ \ D_{\rho,x} > 1/2 \;. \end{matrix} \right.
\end{align}
The only difference with the QSV protocol of Sec.~\ref{sec:QSV} is that some states with $D_{\rho,x}\le 1/8$ can be rejected by $\cE$ with probability greater than 1/3.
This in itself can be an issue for variational approaches as they will be rejecting many useful states in the optimization loop. Nevertheless,
 the ideas behind Thms.~\ref{thm:main} and~\ref{thm:PM} provide similar results for these
 weaker QSV procedures:
\begin{theorem}
\label{thm:weak}
Consider any instance of the QLSP, specified by $A$ and $\vec b$,
and any protocol for QSV as above. Then, for all quantum states $\rho$ that satisfy $C(\rho) \le C_{\min}$, the number 
of $cU_b^{\pm 1}$'s required to implement $\cE$ on input $\sigma_0=\rho^{\otimes m}$ satisfies
\begin{align}
\label{eq:weak}
{\rm Pr} \left( q_{A,b} >  \frac 1 {13} \; \frac \kappa { \|A^{-1} \ket b\| } \right) \ge \frac{1}{6} \; .
\end{align}
\end{theorem}

The proof of Thm.~\ref{thm:weak} follows exactly the same steps as the proof of Thm.~\ref{thm:main} given in Appendix~\ref{app:mainproof}, except that in Eqs.~\eqref{eq:appmainproof1}--\eqref{eq:ineq1} we consider states $\sigma_0$
that are $m$ copies of a state $\rho$, where $C (\rho) \le C_{\min}$.

\vspace{3pt}

Therefore, the complexity of known variational approaches to the QLSP, as measured by the number of uses of $cU_b^{\pm 1}$ or number of copies of $\ket b$ required for their implementation, will be large in worst-case and typical instances, scaling polynomially in $\kappa$. For example, one can use the expectation value of
\begin{align}
\label{eq:Hdef}
    H = A U_b^{\;} P_0^\perp U_b^{-1} A 
\end{align}
as the cost function, where $P_0^\perp = \one -\ketbra 0$ is the projector orthogonal to $\ket 0$. 
This Hamiltonian is positive semi-definite and $\sket{x}$ is its unique ground state with zero eigenvalue~\cite{SSO19}. In Appendix~\ref{app:Hgap} we show that the spectral gap of $H$ is $\cO(1/\kappa^2)$ if the eigenvalue of $A$ with the second smallest magnitude is $\cO(1/\kappa)$, which will be the case in most instances when $N\gg \kappa$.
Determining if $D_{\rho,x}\le 1/8$ for these cases then requires measuring $C$ within additive accuracy that is also $\cO(1/\kappa^2)$; that is, $C_{\min}=\cO(1/\kappa^2)$. 
Because of sampling noise, the overall number of state preparations, projective measurements, and uses of $cU_b^{\pm 1}$ (or $U_b^{\pm 1}$) needed is $\Omega(\kappa^4)$ to obtain the desired accuracy, and this grows rapidly with $\kappa$. Other cost functions may suffer from similar complications~\cite{AL19,HBR19,BLC19}.
 
A version of Thm.~\ref{thm:PM} for the weaker form of QSV discussed above can also be proven.
In this case, Eq.~\eqref{eq:weak} applies under the additional  condition $C(\rho) \le C_{\min}$.

 We note that our lower bounds
 can be bypassed if the structure of $U_b$
 (or $\ket b$) can be exploited, opening the possibility
 to novel quantum approaches for the QLSP that work in this scenario. 
Additionally, the lower bounds are polynomial in $\kappa$ for worst-case and typical instances but, for instances where, e.g., $\|A^{-1} \ket b\|=\Omega(\kappa)$, the number of uses of $cU_b^{\pm 1}$ or copies of $\ket b$ for QSV is constant.
 In this best-case scenario, the number of uses of $cU_b^{\pm 1}$ needed by the quantum algorithms for the QLSP in Refs.~\cite{HHL09,Amb12,CKS17} is also a constant but the query complexity (uses of $U_A$) is still polynomial in $\kappa$,
 while the complexity of variational approaches  remains unknown
 in general.
 
 Other versions of the QLSP 
 for which the goal is to obtain specific
 properties of $\ket x$, rather than preparing $\ket x$,
 may also be of interest.  Our lower bounds do not necessarily apply
 to such versions. For example,
 computing the expectation $\bra x O \ket x$, for some operator $O$, is equivalent to computing $\bra b A^{-1} O A^{-1}\ket b/\|A^{-1} \ket b\|^2$. There are many ways to determine $\bra x O \ket x$ from expectations in $\ket b$ alone, without requiring the preparation or verification of $\ket x$.

\section{Conclusions}
\label{sec:conc}

We studied the complexity of QSV in the context of solving systems of linear equations. We showed that, for worst-case and typical instances of the QLSP, QSV requires a number of state preparation unitaries, and their inverses, that is polynomial in $\kappa$.
This complexity is large for many applications~\cite{Ede88}.

Our results place limitations for approaches to the QLSP that require a verification step (e.g., known variational approaches), where state preparation, gate, or measurement errors will need to decrease fast with $\kappa$ for these instances, if no quantum error correction is used. We note, however, that our results assume no knowledge on the inner workings of the state preparation unitaries. If such knowledge is provided, it may be exploited for more efficient QSV and for solving the QLSP faster.

Our formulation of the QSV problem is fairly generic and concerns the non-adversarial scenario in the sense of Ref.~\cite{ZH19}. Nevertheless, extensions of our results to the adversarial case, in which the input is not promised to be $m\ge 1$ copies of a state $\rho$, would be interesting.
Many quantum operations can be used for QSV,
including those that solve the QLSP or provide estimates of
various distance measures between quantum states,
such as the fidelity. As an example, we provided an optimal QSV procedure based on the HHL algorithm. 

We also discussed a number of open problems
that aim at bypassing our lower bounds. These include
analyzing the complexity of QSV and algorithms for the QLSP in best-case instances, and relaxations of the QLSP
where only certain properties of the quantum state need to be reproduced. Our lower bounds for worst-case and typical instances do not apply to these cases, opening the possibility to faster quantum algorithms.

\section{Acknowledgements}

We thank P. Coles, L. Cincio, and T. Volkoff for discussions. 
This work was supported by the Laboratory Directed Research and Development program of Los Alamos National Laboratory and by the US Department of Energy, Office of Science, Office of Advanced Scientific Computing Research, Quantum Algorithms Teams, 
Accelerated Research in Quantum Computing,  
and the Quantum Computing Application Teams programs. This work is also supported by the Quantum Science Center (QSC), a National Quantum Information Science Research Center of the U.S. Department of Energy (DOE).
YS acknowledges support from the LANL ASC Beyond Moore’s Law project.
Los Alamos National Laboratory is managed by Triad National Security, LLC, for the National Nuclear Security Administration of the US Department of Energy under Contract No. 89233218CNA000001.

\newpage
              
\vspace{1.0cm}
\bibliography{QuantumStateVerification}

\clearpage
\onecolumngrid
\appendix

\section{Proof of Thm.~\ref{thm:main}} 
\label{app:mainproof}

We consider any QSV protocol that, for (any) input $A$ and $\vec b$, provides a quantum operation $\cE$ that satisfies Eq.~\eqref{eq:QSVcond}. As explained, $\cE$ can be described as in Fig.~\ref{fig:quantumprocess}. We let $q$
be the total number of relevant unitaries, that is, the number of $cU_b^{\pm 1}$'s needed to
describe this operation.
Note, however, that the actual number of such unitaries needed on any one execution of the operation and on any one instance of the QLSP, $q_{A,b}$, may be random and less than $q$. For example, the operation can stop
after certain number of amplitude amplification rounds (this would be the case if we use the HHL algorithm for QSV) or after a certain measurement outcome. Then, without loss of generality, each $\cE_j$ in Fig.~\ref{fig:quantumprocess} outputs a (random) bit $s$ that indicates whether 
a stopping criteria has been reached ($s=1$) or not ($s=0$).
Once $s$ is in the state $1$ -- say after the execution of $\cE_j$ -- the remaining operations $\cE_{j+1}, \ldots, \cE_{q+1}$ act trivially and do not alter the state; such  operations are controlled on the state of $s$.

The proof of the lower bound considers
a pair of instances of the QLSP specified by some fixed $A$ and vectors $\vec b$ and $\vec {b'}$, but such that $\ket x$ and $\sket {x'}$ satisfy $D_{x,x'}:=\frac 1 2 \|\ketbra {x'} - \ketbra x \|_\tr>5/8$. Here, $\sket{x'}$ is the solution to the QLSP for initial state $\sket{b'}=U_{b'}  \ket 0$, i.e. $\sket {x'} := A^{-1} \sket{b'}/\|A^{-1} \sket{b'}\|$. We write $\cE$ and $\cE'$
for the quantum operations output by the QSV protocol in either instance. These operations use the controlled unitaries $cU_b^{\pm 1}$ and $cU_{b'}^{\pm 1}$, respectively, at most $q$ times. Note that $\cE'$ can be obtained from $\cE$ by replacing $cU_b^{\pm 1} \rightarrow cU_{b'}^{\pm 1}$.

According to Eq.~\eqref{eq:QSVcond}, $\cE$ must accept a state $\rho$ that satisfies $D_{\rho,x} \le 1/8$ with high probability ($\ge 2/3$) while $\cE'$ must reject the same state $\rho$ with high probability ($\ge 2/3$) since
\begin{align}
    D_{\rho,x'} &\ge | D_{x,x'}-D_{\rho,x}| \\
    & > 5/8 -1/8 \\
    & = 1/2 \;.
\end{align}

We define
\begin{align}
\label{eq:qb0def}
    q_0:=\left \lfloor  \frac 1 {6 \max_{\ket \psi} \sqrt{1-|\bra \psi cU_b^{-1} cU_{b'}^{\;} \ket \psi |^2} }\right \rfloor\;.
\end{align}
We will first show that, with probability at least $1/6$, more than $q_0$ unitaries $cU_b^{\pm 1}$ are needed to implement $\cE$ when the input state $\sigma_0$ contains $m \ge 1$ copies of a state $\rho$ that satisfies $D_{\rho,x}\le 1/8$. Our proof is by contradiction. Let us assume that, with probability $P>5/6$, the operation $\cE$ requires $q_{A,b} \le q_0$ unitaries in this input. On the one hand, in order to satisfy Eq.~\eqref{eq:QSVcond}, the probability of $s=1$ and $r=1$ after this many uses of $cU_b^{\pm 1}$  must be larger than $1/2$.
 This follows from the observation that such probability takes its minimum value ($>1/2$) if the QSV procedure outputs always $r=1$,
in that input, when more than $q_0$ unitaries are used.
On the other hand, the probability of $s=1$ and $r=1$ after $q_{A,b}\le q_0$ uses of $cU_{b'}^{\pm1}$ must be at most $\frac{1}{3}$ for the operation $\mathcal{E}'$ acting on the input state $\sigma_0$ in order to satisfy Eq.(3). This follows from the observation that such probability takes its maximum value if the QSV procedure always outputs $r=0$ when more than $q_0$ unitaries are used. 

We consider the action of $\cE$ and $\cE'$ on the same input state $\sigma_0$. The states produced by these operations after $q_0$ uses of $cU_{b}^{\pm 1}$ and $cU_{b'}^{\pm 1}$ satisfy
\begin{align}
\label{eq:appmainproof1}
\frac 1 2 \|\cE_{q_0} \circ\cdots \circ \cE_1 (\sigma_0) - \cE'_{q_0} \circ\cdots \circ \cE'_1( \sigma_0 ) \|_{\tr} & = \frac 1 2 \|\sum_{j=1}^{q_0} \cE'_{q_0}\circ \cdots \circ \cE'_{j+1} \circ (\cE_j - \cE'_j) \circ \cE_{j-1} \circ \cdots \circ \cE_1(\sigma_0) \|_{\tr} \\
    & \le \frac 1 2 \sum_{j=1}^{q_0} \|  (\cE_j-\cE'_j)(\sigma_{j-1})  \|_{\tr} \\
    & = \frac 1 2 \sum_{j=1}^{q_0} \| (\cU_b^{s_j} - \cU_{b'}^{s_j}) \circ {\cal F}_j (\sigma_{j-1})\|_{\tr}  \\
    & \le  q_0 \|\cU_b - \cU_{b'}\|_{\diamond} \\
    \label{eq:ineq0}
    & =   q_0  \max_{\ket \psi} \sqrt{1-|\bra \psi cU_b^{-1} cU_{b'}^{\;} \ket \psi |^2} \\
    \label{eq:ineq1}
    & \le 1/6\; .
\end{align}
The state $\sigma_j$ is  obtained after the action of $\cE_j \circ \cdots \circ \cE_1$ on $\sigma_0$ and
$\cU_b^{s_j}$ and $\cU_{b'}^{s_j}$ are the quantum operations
that implement the unitaries $cU_b^{s_j}$ and $cU_{b'}^{s_j}$, respectively  ($s_j=\pm 1$).
The diamond norm of two channels $\cE$ and $\cE'$ is defined in the standard way as $\|\cE-\cE'\|_\diamond=\max _\tau \frac 1 2 \|\cI \otimes \cE(\tau) -\cI \otimes \cE'(\tau)\|_\tr$, where $\cI$ is a trivial operation acting on a different subsystem and $\tau$ is the state of the composite system.
Note that
$\| \cU_b - \cU_{b'} \|_{\diamond} = \| \cU_b^{-1} - \cU_{b'}^{-1} \|_{\diamond}$. 
Equation~\eqref{eq:ineq0} follows directly from
the property $\frac 1 2 \|\ketbra {\psi_1} - \ketbra {\psi_2 }\|_{\tr}= \sqrt{1-|\langle \psi_1 | \psi_2 \rangle|^2}$, where $\ket{\psi_1}$ and $\ket{\psi_2}$ are any two unit states,
and Eq.~\eqref{eq:ineq1} follows from Eq.~\eqref{eq:qb0def}.

Therefore, using the operational
meaning of the trace distance,
$\cE'$ would accept $\sigma_0$ using $q_{A,b} \le q_0$
unitaries $cU_{b'}^{\pm 1}$, with probability larger than $1/2- 1/6=1/3$.
But this contradicts Eq.~\eqref{eq:QSVcond}.
It follows that the probability that $\cE$ 
uses $q_{A,b} \le q_0$ unitaries
$cU_{b}^{\pm 1}$ satisfies $P \le 5/6$. Equivalently, the probability that $\cE$ uses more than $q_0$ such unitaries in this input
is, at least, $1/6$.

\subsubsection{Pairs of instances}
\label{app:pair}

\begin{figure}[htb]
    \includegraphics[width=11cm]{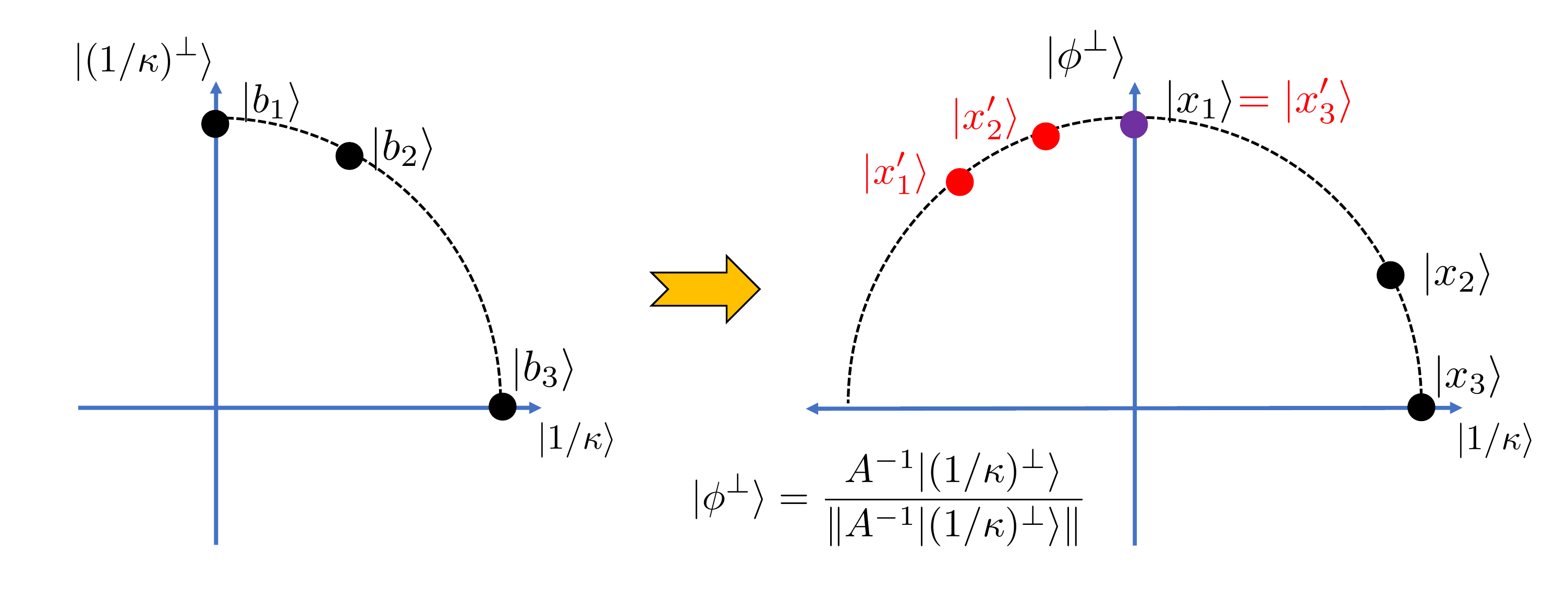}
     \caption{Geometric representation of three pairs of instances used to prove Thm.~\ref{thm:main}, assuming that $A$ has an eigenvalue $1/\kappa$.}
    \label{fig:Theorem1}
\end{figure}

For every instance of the QLSP specified by an $A$ and $\vec b$, we construct another one that satisfies the assumptions of the previous analysis and will provide the lower bound in Thm~\ref{thm:main}. We assume that $A$ has an eigenvalue $1/\kappa$ but,
if $A$ has an eigenvalue $-1/\kappa$ instead, a simple modification in the following proof (a redefinition of $\sket{\tilde b}$ below) provides the same result.
We write
\begin{align}
    \ket b = v \ket{( 1/\kappa)} + v^\perp \sket{( 1/\kappa)^\perp} \;,
\end{align}
where the unit state $\sket{( 1/\kappa)}$ is an eigenstate of $A$
of eigenvalue $ 1/\kappa$, $\sket{( 1/\kappa)^\perp}$
is a unit state orthogonal to $\sket{ (1/\kappa)}$, and $v \ge 0$, $v^\perp \ge 0$, $v^2+(v^\perp)^2=1$. In case $\sket{b} = \sket{( 1/\kappa)}$, i.e. $v=1$, $\sket{( 1/\kappa)^\perp}$ can be any unit state orthogonal to $\sket{ (1/\kappa)}$. We define
\begin{align}
    \sket{\phi^\perp}:= \frac{A^{-1} \sket{( 1/\kappa)^\perp}} {\| A^{-1} \sket{( 1/\kappa)^\perp}\|} \;,
\end{align}
which is also a unit state orthogonal to 
$\sket{( 1/\kappa)}$. Then,
\begin{align}
    \ket x \propto  \kappa v \sket{( 1/\kappa)} + \| A^{-1} \sket{( 1/\kappa)^\perp}\| v^\perp \sket{\phi^\perp} \; ,
\end{align}
and we note that 
\begin{align}
    1 \le \| A^{-1} \sket{( 1/\kappa)^\perp}\| \le \kappa \;.
\end{align}
The other instance is defined such that
\begin{align}
 \sket{b'}=\frac{\sket{\tilde b}}{\| \sket{\tilde b}\|} \; , \; \sket{\tilde b} := 
 &  \ket{b} + \frac {\|A^{-1}\ket b\|} \kappa  (- \sket{( 1/\kappa)} + (1/5) \sket{( 1/\kappa)^\perp} ) \; .
\end{align}
With this choice, we obtain $\sket{b'} \ne 0$ and
\begin{align}
\label{eq:x'}
    \sket{x'} \propto \ket x - \sket{( 1/\kappa)} + 
    \frac {\|A^{-1}\sket {( 1/\kappa)^\perp}\|} {5 \kappa} \sket{\phi^\perp} \;.
\end{align}
We give a geometric representation of pairs of these  instances in Fig.~\ref{fig:Theorem1}, pictured in the corresponding two-dimensional subspaces.

The initial states for the corresponding two instances satisfy
\begin{align}
    \| \sket{b}-\sket {b'}\| & \le  \| \sket b- \sket{\tilde b} \|+ \| \sket{\tilde b} - \sket{b'} \| \\
    &=  \| \sket b- \sket{\tilde b} \| + \| ( \| \sket{\tilde b}\|-1) \sket{b'} \| \\
    &=  \| \sket b- \sket{\tilde b} \| + |  \| \sket{\tilde b}\|-1  | \\
    & =  \| \sket b- \sket{\tilde b} \|+ | \| \sket{\tilde b}\| -\|\ket b\| |  \\
    &\le 2 \| \sket{\tilde b} - \ket b   \|  \\
    & \le \frac{2\sqrt {26}}{5} \frac{ \|A^{-1} \ket b\|} \kappa \; .
\end{align}
Then, there exist two unitaries $U_b^{\;}$ and $U_{b'}^{\;}$
that prepare the states $\ket b$ and $\sket{b'}$, respectively, and satisfy
\begin{align}
\label{eq:U_bbound}
\| U^{\;}_{b} - U^{\;}_{b'} \|& =\|cU^{\;}_b - cU^{\;}_{b'}\|
     \\
    & = \| \sket{b}-\sket {b'}\| \\
    \label{eq:normofbminusbprime}
    & \le \frac{2\sqrt {26}}  5 \frac{ \|A^{-1} \ket b\| } \kappa\; .
\end{align}
These unitaries can be explicitly constructed in many ways;
for example, $U_{b'}$ can be $U_b$ followed by a rotation in the two-dimensional subspace, along an axis that is orthogonal to the plane formed by $\ket b$ and $\sket{b'}$, that takes $\ket b$ to $\sket{b'}$:
\begin{align}
    U_{b'}& = e^{i \theta M }U_b \; , \\
    \label{eq:costheta}
    \theta &= \arccos (\langle b' | b \rangle ) \;, \\
    M & = i \ket b \sbra{b^\perp} - i \sket{b^\perp} \bra b \;, \\
    \sket{b^\perp} &= (\one - \ketbra b) \sket{b'}/ \| (\one - \ketbra b) \sket{b'} \| \;.
\end{align}
Note that $1 \ge \langle b' | b \rangle \ge 0$ so that $\pi/2 \ge \theta \ge 0$.
Additionally, these unitaries satisfy
\begin{align}
\max_{\ket \psi} \sqrt{1-|\bra \psi cU_b^{-1} cU_{b'}^{\;} \ket \psi|^2} & = \max_{\ket \psi} \sqrt{1-|\bra \psi \ketbra 0 \otimes \one + \ketbra 1 \otimes U_b^{-1} U_{b'}^{\;} \ket \psi|^2} \\
& =\max_{\ket \psi} \sqrt{1-|\bra \psi \ketbra 0 \otimes \one + \ketbra 1 \otimes e^{i \theta M} \ket \psi|^2} \\
& \le \max_{\ket \psi} \sqrt{1-|\bra \psi (\ketbra 0 + \cos \theta \ketbra 1) \otimes \one  \ket \psi|^2} \\
& \le \sqrt{1-(\cos \theta)^2}\\
& = \sin \theta \\
& \le \| \ket b -\sket{b'}\| \\
    \label{eq:U_bboundb}
    &\le \frac{2\sqrt {26}}  5 \frac{ \|A^{-1} \ket b\| } \kappa\; .
\end{align}

The state $\sket{x'}$
in Eq.~\eqref{eq:x'} is closest to $\ket x$ when $\|A^{-1} \sket{(1/\kappa)^\perp}\|=\kappa$ and $\ket x=\sket{\phi^\perp}$, as can be observed from Fig.~\ref{fig:Theorem1}. In this case
$\sket{x'}=\sqrt{1/61}(6 \sket{\phi^\perp}-5\sket{( 1/\kappa)})$
and $|\langle x |x'\rangle |=6/\sqrt{61}$. Then, in general,
\begin{align}
    D_{x',x}& =\frac 1 2 \|\ketbra {x'} - \ketbra x \|_\tr \\
    & = \sqrt{1 - |\langle x | x' \rangle|^2} \\
    & \ge \sqrt{25/61} \\
    & > 5/8 \;.
\end{align}

Thus, these pairs of instances satisfy the assumptions
of the previous analysis.
Equations~\eqref{eq:qb0def} and~\eqref{eq:U_bboundb}
imply that, with probability at least $1/6$, more than
\begin{align}
    q_0& \ge \left \lfloor \frac {5} {12 \sqrt{26}} \frac \kappa {\| A^{-1} \ket b\|} \right \rfloor \\
    & \ge \left \lfloor\frac 1 {13} \frac \kappa {\| A^{-1} \ket b\|} \right \rfloor \; 
\end{align}
unitaries $cU_b^{\pm 1}$ are required to implement $\cE$
when the input state contains $m$ copies of $\rho$, and $D_{\rho,x} \le 1/8$. $\qed$

\section{Proof of Thm.~\ref{thm:typical}}
\label{app:typical}

Let $\sket{\bar b}=\sum_{\lambda} a_\lambda \ket \lambda$ be a quantum state proportional to $\ket b$, i.e. $\ket b:=\sket{\bar b}/\|\sket{\bar b}\|$, and $\ket \lambda$ be an eigenvector of $A$ of eigenvalue $\lambda$. In particular, the amplitudes of $\sket{\bar b}$ satisfy
\begin{align}
    \Pr(p_\lambda)=N e^{-N p_\lambda} \;,
\end{align}
where $p_\lambda=|a_\lambda|^2$ and
$\Pr(p_\lambda)$ is the Porter-Thomas distribution.
Standard probability rules imply
\begin{align}
\Pr \left\{ \|A^{-1}\sket{ b} \| \notin[\sqrt{\kappa/2}, \sqrt{3  \kappa/2}]\right\} & = \Pr \left\{\frac{\|A^{-1}\sket{\bar b} \|}{\|\sket{\bar b}\|} \notin[\sqrt{\kappa/2}, \sqrt{3  \kappa/2}]\right\} \\
& =
\Pr \left\{\left(\frac{\|A^{-1}\sket{\bar b} \|}{\|\sket{\bar b}\|} \right)^2 \notin[ \kappa/2, 3  \kappa/2]\right\} \\
\label{eq:prob_inequalities}
&\leq \Pr \left\{\|\sket{\bar b}\|^{2} \notin[5/6, 3/2]\right\} + \Pr \left\{\|A^{-1} \sket{\bar b}\|^{2} \notin[ 3 \kappa /4, 5 \kappa / 4]\right\} \;.
\end{align}
We will upper bound each term of Eq.~\eqref{eq:prob_inequalities} below.

First we focus on $\|\sket{\bar b}\|^2 = \sum_\lambda p_\lambda$.
We apply Chernoff's bound twice; once for establishing an upper bound on the probability that $\|\sket{\bar b}\|^2 \ge 3/2$ and then on the probability that $\|\sket{\bar b}\|^2 \leq 5/6$.
 Since the $p_\lambda$'s are i.i.d., we obtain
\begin{align}
\label{eq:chernoff_upper}
\operatorname{Pr}\{\|\sket{\bar b}\|^2 &\geq 3/2\} \leq  \min _{t>0} e^{-3 tN /2} \left({\rm E}\left[e^{tN p_{\lambda}}\right]\right)^{N} \; , \\
\label{eq:chernoff_lower}
\operatorname{Pr}\{\|\sket{ \bar b}\|^2 &\leq 5/6) \} \leq \min _{t>0} e^{5tN /6} \left({\rm E}\left[e^{-tN p_{\lambda}}\right]\right)^{N} \; ,
\end{align}
where ${\rm E}[\cdot]$ is the expectation value. For the Porter-Thomas distribution and $0<t<1$,
\begin{align}
    {\rm E}[e^{t N p_\lambda}] &= \frac{1}{1- t} \; ,
\end{align}
and for $t>0$,
\begin{align}
    {\rm E}[e^{-t N p_\lambda}] &= \frac{1}{1+ t} \; .
\end{align}
The minimization over $t$ in Eqs.~\eqref{eq:chernoff_upper} and~\eqref{eq:chernoff_lower} can be performed analytically. Nevertheless, we can pick a suitable value for $0< t<1$ such that the upper bounds decay exponentially with $N$. In particular, for $t=1/4$, we obtain
\begin{align}
\label{eq:chernoff_upper2}
\Pr\{\|\sket{\bar b}\|^2 \geq 3/2\} & \leq  e^{-3N /8} (4/3)^{N} \\
& \le e^{-0.087 N}\; , 
\end{align}
and
\begin{align}
\label{eq:chernoff_lower2}
\Pr \{\|\sket{ \bar b}\|^2 \leq 5/6) \} &\leq e^{5 N /24} (4/5)^{N} \\
& \le e^{-0.014N}\; .
\end{align}

Next we focus on $\|A^{-1} \sket{\bar b}\|^2 = \sum_\lambda p_\lambda/\lambda^2$ and assume that the eigenvalues are sampled from $\rm{unif}\left \{ [-1,-1/\kappa] \cup [1/\kappa,1] \right \}$, that is, the uniform distribution in $[-1,-1/\kappa] \cup [1/\kappa,1]$. 
Chernoff's bound implies
\begin{align}
\label{eq:chernoff_upper3}
\Pr\{\|A^{-1} \sket{\bar b}\|^2 &\geq 5 \kappa/4\} \leq  \min _{t>0} e^{-5 tN \kappa/4} \left({\rm E}\left[e^{tN p_{\lambda}/\lambda^2}\right]\right)^{N} \; , \\
\label{eq:chernoff_lower3}
\Pr\{\|A^{-1} \sket{\bar b}\|^2 &\leq 3\kappa/4) \} \leq \min _{t>0} e^{3tN \kappa/4} \left({\rm E}\left[e^{-tN p_{\lambda}/\lambda^2}\right]\right)^{N} \; .
\end{align}
Again, we can perform the minimization in $t$ but it suffices to pick a suitable $t$ that provides useful, exponentially decaying bounds. In particular, for $t = 1/(8\kappa^2)$, 
\begin{align}
   {\rm E}\left[e^{tN p_{\lambda}/\lambda^2}\right]&=  \frac 1 {1-1/\kappa} \int_{1/\kappa}^1 d\lambda \int_0^\infty dp_{\lambda} \; N e^{-Np_{\lambda}} e^{tNp_{\lambda}/\lambda^2}  \\
   & = \frac 1 {1-1/\kappa} \int_{1/\kappa}^1 d\lambda \; \frac 1 {1- t /\lambda^2} \\
   & \le \frac 1 {1-1/\kappa} \int_{1/\kappa}^1 d\lambda  \; (1+ (8/7) t /\lambda^2) \\
   & = 1 + 8t \kappa/7 \\
   & \le e^{8 t \kappa/7} \\
   & = e^{1/(7\kappa)}\;,
\end{align}
and
\begin{align}
  {\rm E}\left[e^{-tN p_{\lambda}/\lambda^2}\right]&=  \frac 1 {1-1/\kappa} \int_{1/\kappa}^1 d\lambda \int_0^\infty dp_{\lambda} \; N e^{-Np_{\lambda}} e^{-tNp_{\lambda}/\lambda^2}  \\
   & = \frac 1 {1-1/\kappa} \int_{1/\kappa}^1 d\lambda \; \frac 1 {1+ t /\lambda^2} \\
   & \le \frac 1 {1-1/\kappa} \int_{1/\kappa}^1 d\lambda \; (1- (8/9) t/\lambda^2) \\
   & = 1 - 8 t \kappa /9 \\
   & \le e^{-8 t \kappa/9} \\
   & = e^{-1 /(9 \kappa)}\; .
\end{align}
Using these bounds in Eqs.~\eqref{eq:chernoff_upper3} and~\eqref{eq:chernoff_lower3} gives
\begin{align}
\label{eq:chernoff_upper4}
\Pr\{\|A^{-1} \sket{\bar b}\|^2 &\geq 5 \kappa/4\} \leq  e^{-0.013 N/\kappa} \; , \\
\label{eq:chernoff_lower4}
\Pr\{\|A^{-1} \sket{\bar b}\|^2 &\leq 3\kappa/4 \} \leq e^{-0.017 N /\kappa} \; .
\end{align}
Last, since $\kappa \ge 1$,  the right hand side of Eq.~\eqref{eq:prob_inequalities} can be upper bounded by
\begin{align}
    e^{-0.087N} + e^{-0.014N} + e^{-0.013 N/\kappa}+ e^{-0.017 N /\kappa} \le 
    4 e^{-0.013 N/\kappa} \;.
\end{align}
\qed

\section{QSV via the HHL algorithm}
\label{app:HHLQSV}

We construct an
operation for QSV that uses a number of $cU_{b}^{\pm 1}$'s that is almost optimal. This operation first solves the QLSP and then compares the outcome to the state one wishes to verify. Several algorithms can  be used for solving the QLSP but, for simplicity, we  consider the HHL algorithm here.  
The HHL algorithm is not optimal for solving the QLSP in terms of its scaling with respect to $\kappa$ or a precision parameter~\cite{Amb12,CKS17,SSO19}. However, it turns out that it can be used in a QSV algorithm that is  almost optimal in terms of the uses of $cU_b^{\pm1}$'s because for this purpose we only need to prepare states that are within a constant distance from $\ket{x}$.

A key subroutine of the HHL algorithm is based on quantum phase estimation and implements a conditional rotation on an ancillary qubit as follows. 
Let $\ket{b}=\sum_\lambda b_\lambda \sket{\lambda}$ and $\|\ket b\|=1$. If we ignore errors for the moment, 
this subroutine implements a unitary $U$ such that
\begin{align}
    U \ket{b}\ket 0 \ket{0\ldots 0} & \rightarrow \sum_\lambda b_\lambda \sket{\lambda} \left(  \frac{1}{\kappa \lambda} \ket{0} + \sqrt{1-\frac{1}{\kappa^2 \lambda^2}}\ket{1}  \right) \ket {0 \ldots 0} \\
    \label{eq:HHLexact}
    & = \left(   \frac {A^{-1}}{\kappa}   \ket b \right) \ket 0 \ket{0 \ldots 0} + (h(A)\ket b) \ket 1 \ket{0 \ldots 0}\; ,
\end{align}
where $h(A)$ does not implement the matrix inversion. The unitary
$U$ depends on $A$ and implements other two-qubit gates (e.g., for the quantum Fourier transform) but does not use $cU_b^{\pm 1}$ (or $U_b^{\pm 1}$). If the ancillary qubit is measured and the outcome is $\ket 0$, then the first register is exactly in the  state $\ket{x}$. This occurs with probability
\begin{align}
    p_\text{success} &= \sum_\lambda \frac{\vert b_\lambda\vert^2}{\kappa^2 \lambda^2} \\
    & = \left( \frac {\| A^{-1} \ket{b}\| } \kappa \right)^2 \;.
\end{align}
Rather than measuring this qubit, one can implement
amplitude amplification to boost the probability of
measuring this qubit in $\ket 0$ and preparing $\ket x$ to almost 1.
This approach would require $ \cO(1/\sqrt{p_\text{success}})$
reflections over the state $U \ket{b}\ket 0 \ket{0\ldots 0}$ in expectation,
which translates to $\cO(\kappa/\|A^{-1}\ket b\|)$ uses of $cU_b^{\pm 1}$ in expectation, following Ref.~\cite{BHMT02}. 

Once the state $\ket x$ is prepared, we can perform QSV via the swap or Hadamard test~\cite{BCW01}.
Using one copy of  $\rho$ and one copy of $\ket x$,
the swap test performs a joint operation and outputs a bit $r'$ that satisfies
$\Pr(r'=1)= (1+\bra x \rho \ket x)/2$. 
If $D_{\rho,x}\le 1/8$, then $\bra x \rho \ket x\ge 1-D_{\rho,x} \ge 1-1/8$ and $\Pr(r'=1)\ge 15/16$.
If $D_{\rho,x} > 1/2$, then $\bra x \rho \ket x \le 1- D_{\rho,x}^2< 1-1/4$ and $\Pr(r'=1) <7/8$.
To produce the bit $r$ with the desired properties
of Eq.~\eqref{eq:QSVcond}, we can implement the swap test and sample $r'$, say, 64 times. Let $r=1$ only when the Hamming weight of the string is 59 or more, and $r=0$ otherwise. Then,
\begin{align}
\label{eq:HHLr=1prob}
    \Pr(r=1)=\sum_{k=0}^5 \frac{64!}{k! (64-k)!} (1-\Pr(r'=1))^k(\Pr(r'=1))^{64-k} \;,
\end{align}
and if 
$D_{\rho,x} \le 1/8$ or $D_{\rho,x} > 1/2$, we obtain $\Pr(r=1) \ge 0.79 > 2/3$ or $\Pr(r=1)<.18 < 1/3$, respectively.
As the swap test is used a constant number of times, the above QSV procedure can be implemented using the HHL algorithm a constant number of times or, equivalently, using the unitaries  $cU_b^{\pm 1}$ a number of times that is $\cO(\kappa/\|A^{-1}\ket b\|)$ in expectation.

\subsubsection{Effects of errors}

The previous analysis would suffice to prove that the HHL algorithm is optimal for QSV if $U$ could be implemented exactly. However, due to imprecise quantum phase estimation, the HHL algorithm implements a unitary $\tilde U$ that approximates the transformation in Eq.~\eqref{eq:HHLexact}. Once the ancillary qubits are discarded, the quantum state prepared by HHL is $\rho_x$ and satisfies
\begin{align}
    D_{\rho_x,x} \le \epsilon \;,
\end{align}
for arbitrary $\epsilon >0$.
While the actual value of $\epsilon$ may not  affect the number of $cU_b^{\pm 1}$'s needed to implement the QSV procedure, we will show that a constant $\epsilon$ suffices.

Let $\epsilon \le 1/100$. 
Then, when the input to the swap test is one copy of $\rho$ and one copy of $\rho_x$, the test produces a bit $r'$ satisfying $\Pr(r'=1) \ge 15/16-1/100$ if $D_{\rho,x}\le 1/8$ and $\Pr(r'=1)< 7/8+1/100$ if
$D_{\rho,x}> 1/2$. That is, the probabilities of the exact case analysis can only be modified by, at most, $\epsilon$. This is due to a property of the trace distance being non increasing under quantum operations (CPTP maps). As before, we can produce the bit $r$ that satisfies Eq.~\eqref{eq:QSVcond}
by sampling $r'$, say, 64 times. Let $r=1$
when the Hamming weight of the string is 59 or more, and $r=0$ otherwise. If we compute Eq.~\eqref{eq:HHLr=1prob}
for this case, we obtain $\Pr(r=1) \ge 0.68 > 2/3$ if $D_{\rho,x}\le 1/8$ and $\Pr(r=1) <.25 < 1/3 $ if $D_{\rho,x}>1/2$.

In Ref.~\cite{HHL09} it was shown that the probability of success in the preparation of $\rho_x$, $\tilde{p}_\text{success}$, satisfies
\begin{align}
    \frac{\vert \tilde{p}_\text{success} - p_\text{success} \vert}{p_\text{success}} =\cO(\epsilon) \;.
\end{align}
This implies ${\tilde p_\text{success}}=\Omega({p_\text{success}})$
so that the overall number of amplitude amplification rounds in the HHL algorithm is $\cO(1/\sqrt{p_\text{success}})$
in expectation. As the HHL algorithm and the swap test are needed a constant number of times, the unitaries $cU_b^{\pm 1}$ are used $\cO(\kappa/\|A^{-1}\ket b\|)$
times in expectation.

\section{Proof of Thm.~\ref{thm:PM}}
\label{app:PMproof}

Let ${\cal L}$ be the quantum operation provided by a QSV protocol of the PM type
on input $A$, as described in Fig.~\ref{fig:quantumprocess2}. We let $q$ be the  number of copies of $\ket b$ in the input state of $\cal L$. However, the actual number of such states used on any one execution of ${\cal L}$, $ q_{A,b}$, may be random and less than $ q$. As in Appendix~\ref{app:mainproof}, we can assume, without loss of generality, that each ${\cal F}_j$ outputs a bit $s$ that indicates whether a stopping criteria has been reached ($s=1$) or not ($s=0$). Once $s$ is in state 1 -- say after the execution of ${\cal L}_j$ (or ${\cal F}_j)$ -- the remaining operations ${\cal F}_{j+1},\ldots,{\cal F}_{ q}$ act trivially, do not require the preparation of further copies of $\ket b$, and do not alter the state; such operations are controlled on the state of $s$.

The proof  closely follows that of Thm.~\ref{thm:main} given in Appendix~\ref{app:mainproof}. It considers a pair of instances of the QLSP specified by some fixed $A$ and vectors $\vec b$ and $\vec b'$ such that $\ket x$ and $\sket{x'}$ satisfy $D_{x,x'}>5/8$. Then, according to Eq.~\eqref{eq:QSVcond}, the operation ${\cal L}$ that uses copies of $\ket b$ must accept a state $\rho$ that satisfies $D_{\rho,x} \le 1/8$ with high probability ($ \ge 2/3$). The operation ${\cal L}$ that uses copies of $\sket {b'}$ must reject $\rho$ with high probability ($ \ge 2/3$), because this input state corresponds to the other instance.

We define
\begin{align}
\label{eq:qb0def2}
     q_0 : = \left \lfloor \frac 1 {36 (1-|\langle b | b' \rangle |^2)}\right \rfloor \;.
\end{align}
We will first show that, with probability at least $1/6$, more than $q_0$ copies of $\ket b$ are used by ${\cal L}$ when the input state $\sigma_0$ contains $m \ge 1$ copies
of a state $\rho$ that satisfies $D_{\rho,x}\le 1/8$. The proof is also by contradiction. Let us assume that, with probability $P>5/6$, the operation ${\cal L}$ requires $q_{A,b} \le q_0$
copies of $\ket b$. Then, in order to satisfy Eq.~\eqref{eq:QSVcond}, the probability of $s=1$ and $r=1$ 
in the state output by ${\cal L}_{q_{0}}$ must be larger than $1/2$. In addition, as the trace distance is non-increasing under quantum operations (CPTP maps), 
\begin{align}
  \frac  1 2  \|{\cal L}_{q_0}(\rho^{\otimes m}\otimes\ket{b}\!\bra{b}^{\otimes q_0}) - {\cal L}_{q_0}(\rho^{\otimes m}\otimes\ket{b'}\!\bra{b'}^{\otimes q_0}) \|_{\tr}
    & \le D_{\ket b^{\otimes q_0},\sket {b'}^{\otimes q_0} } \\
    & = \sqrt{1-|\langle b | b' \rangle|^{2 q_0}} \\
    & \le \sqrt{ q_0 (1-|\langle b | b' \rangle |^2)} \\
    & \le 1/6 \;.
\end{align}

Therefore, under the assumptions and using the operational meaning of the trace distance, ${\cal L}$ would accept $\rho$
using $ q_{A,b} \le q_0$ copies of $\sket{b'}$ with probability larger than $1/2-1/6=1/3$. But this contradicts Eq.~\eqref{eq:QSVcond}, which states that the probability of $r=1$ should be, at most, $1/3$ in this case. It follows that the probability that ${\cal L}$ requires more than $q_0$ copies of $\ket b$ in this input is lower bounded by $1/6$.

From Eqs.~\eqref{eq:costheta} and~\eqref{eq:normofbminusbprime}, it follows that 
\begin{align}
    1-|\bra{b}\ket{b'}|^2 & = (\sin \theta)^2\\
    & \le \|\ket b - \ket{b'}\|^2 \\
    \label{eq:bboundb}
    & \le \frac {104}{25} \left( \frac{\|A^{-1}\ket{b}\|} \kappa \right)^2 \;.
\end{align}
Equations~\eqref{eq:qb0def2} and~\eqref{eq:bboundb}
imply that, with probability at least $1/6$, more than
\begin{align}
    q_0& \ge \left \lfloor \frac{25}{3744} \left( \frac \kappa {\|A^{-1}\ket{b}\|} \right)^2 \right \rfloor \\
    & \ge \left \lfloor \frac 1 {150} \left( \frac \kappa {\|A^{-1}\ket{b}\|} \right)^2 \right \rfloor
\end{align}
copies of state $\ket{b}$ are required to implement ${\cal L}$
when the input state contains $m$ copies of $\rho$, and $D_{\rho,x} \le 1/8$. $\qed$

\section{Spectral properties of $H$}
\label{app:Hgap}

We analyze some spectral properties of $H=A P_b^\perp A$,
where $P_b^\perp = \one - \ketbra b$ is a projector orthogonal to $\ket b$~\cite{SSO19}. Since $H$
is of the form $B^\dagger B$, then $H \ge 0$ and
\begin{align}
    H \ket x &= A P_b^\perp A \ket x \\
    & = A\frac {P_b^\perp \ket b} {\|A^{-1} \ket b\|} \\
    & = 0 \;.
\end{align}
Moreover, $\ket x$ is the unique eigenstate of eigenvalue 0. 

For any $\sket{x^\perp}$, such that $\bra{x^\perp}\ket{x}=0$, the gap of the Hamiltonian $H$ can be bounded as
\begin{align}
    \Delta &\le \sbra{x^\perp}H\sket{x^\perp} \\
    & = \sbra{x^\perp}A^2\sket{x^\perp}-\vert\sbra{x^\perp}A\ket{b}\vert^
    2 \\
    \label{eq:gap}
    & \le \sbra{x^\perp}A^2\sket{x^\perp}\; .
\end{align}
By assumption, the absolute smallest eigenvalue of $A$ is $1/\kappa$ and we let $\lambda_{ss}$ denote the eigenvalue with the second smallest magnitude. We write 
$\sket{(1/\kappa)}$ and $\sket{\lambda_{ss}}$
for the corresponding eigenstates. 
Without loss of generality 
\begin{align}
    \ket{x} &= a\ket{(1/\kappa)} + b \ket{\lambda_{ss}} + \sqrt{1-a^2-b^2} \ket{\gamma} \; ,
\end{align}
where $\sket{\gamma}$ is a unit state orthogonal to the two-dimensional subspace spanned by $\sket{(1/\kappa)}$ and $\sket{\lambda_{ss}}$. In this subspace, there exists
a unit state $\sket{x^\perp}$ that is orthogonal to $\ket x$, that is, $\langle x^\perp | x \rangle=0$. It satisfies
$\sbra{x^\perp}A^2\sket{x^\perp}\le \lambda_{ss}^2$ and,
together with Eq.~\eqref{eq:gap}, we obtain $\Delta \le \lambda_{ss}^2$.
In particular, this implies $\Delta=\cO(1/\kappa^2)$ whenever $A$ has at least two eigenvalues of magnitude $\cO(1/\kappa)$, which will be the case in most instances when $N \gg \kappa$.

\end{document}